\documentclass[aps,prl,twocolumn,groupedaddress]{revtex4}

\usepackage{graphicx}

\usepackage{wasysym}
\usepackage{color}
\usepackage{amsfonts,amssymb}

\newcommand{\etal}{\textit{et al.\ }}

\begin{document}

\title{Comment on ``Scaling feature of magnetic field induced Kondo-peak splittings''\cite{zhangKondoMag10}}

\author{Sebastian Schmitt and Frithjof B. Anders}
\affiliation{Lehrstuhl f\"ur Theoretische Physik II, Technische Universit\"at Dortmund, Otto-Hahn-Str. 4, 44221 Dortmund, Germany}

\begin{abstract}
In a recent work Zhang and coworkers (PRB 82, 075111 (2010)) studied the Zeeman splitting of the Kondo 
resonance for the single impurity Anderson model in a finite magnetic field $B$ with the numerical 
renormalization group (NRG) method. 
They report a discrepancy between the position of the Kondo resonance in the total spectral function 
and the position in the spin resolved spectral function 
at large magnetic fields.
 Additionally, the position of the Kondo maximum exceeded the 
Zeeman energy for $B/ T_K\gtrsim 5-10$, where $T_K$ is the low energy Kondo scale of the model.
In this comment we argue that both these findings are a result of the specific choice of NRG 
parameter values. However, we reproduce the crossover in the splitting from Kondo-like behavior to 
a non-universal splitting larger than the Zeeman energy, but this crossover occurs at much larger 
fields of the order of the charge scale.
\end{abstract}

\maketitle

In a recent work\cite{zhangKondoMag10} Zhang and coworkers studied the Zeeman splitting of the Kondo
resonance for the single impurity Anderson model  (SIAM) in a finite magnetic field $B$. 
They employed the numerical renormalization
group (NRG) to obtain spectral functions and discussed the position  of the Kondo resonance
in the spin-resolved and total spectral function, 
$\rho_\sigma(\omega)$ and $\rho(\omega)=\sum_\sigma\rho_\sigma(\omega)$, respectively.
Two important points made in the work are:
\begin{enumerate}
\item[(i)]
 With increasing magnetic field $B$ the position $\delta_\uparrow$ of the Kondo 
  resonance in the total spectral function
  $\rho(\omega)$  \textit{does not} approach its position $\Delta_\uparrow$ in the 
  spin resolved spectral function, but instead $\Delta_\uparrow>\delta_\uparrow$.
\item[(ii)]
  The positions $\delta_\uparrow$ and $\Delta_\uparrow$
  exceed the Zeeman energy, i.e.\  $\delta_\uparrow,\Delta_\uparrow > B$ 
  for $B/ T_K\gtrsim 5-10$, where $T_K$ is the low energy Kondo scale 
  of the model ($g=2$, $\mu_B=k_B=\hbar=1$).  
\end{enumerate}
We show in the following, that both these conclusions cannot be drawn from the data presented
in the publication, since they originate from the specific choice of NRG parameters.

\begin{figure}[t]
    \includegraphics[width=8cm]{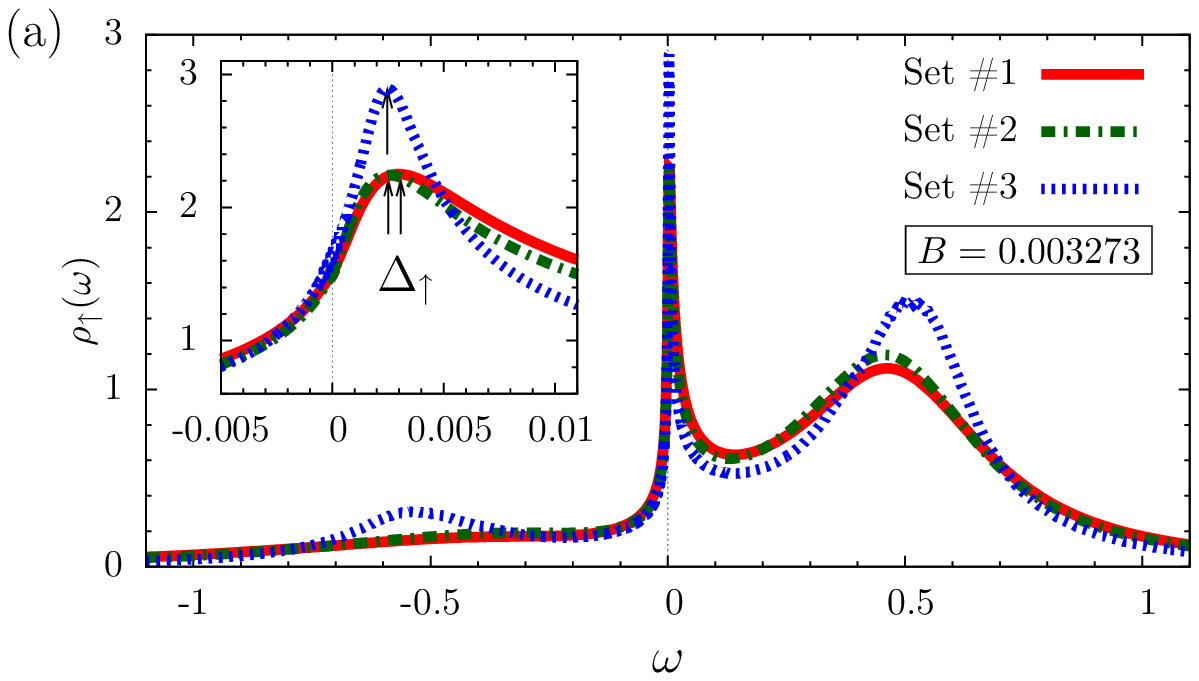}
\includegraphics[width=4cm]{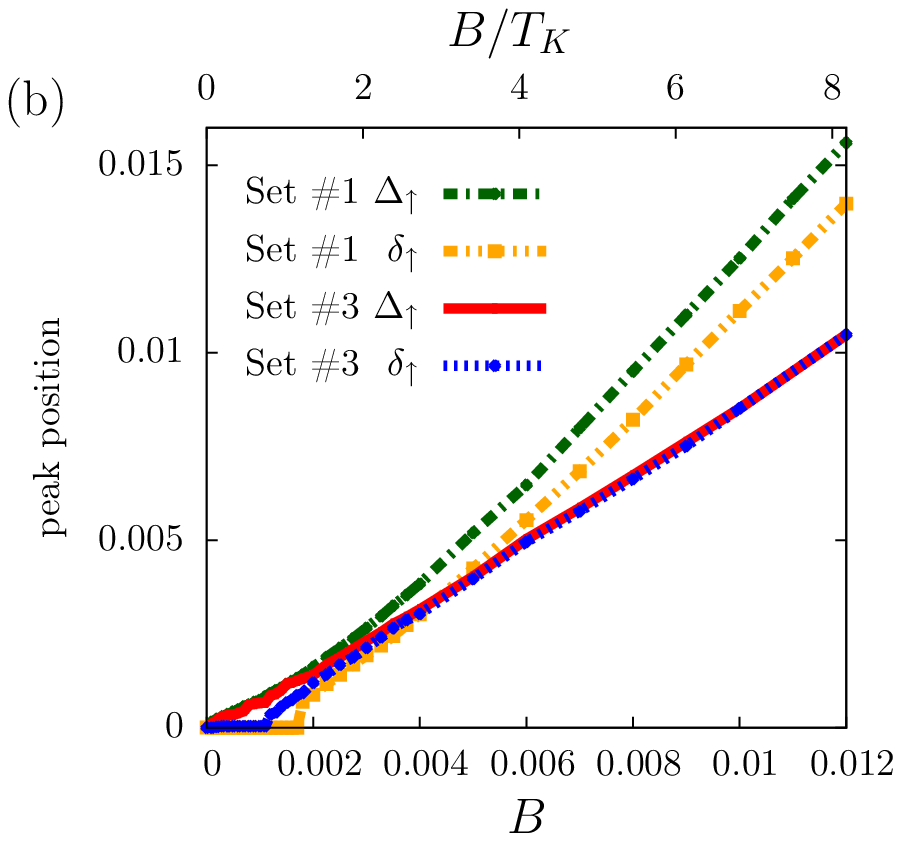}  \includegraphics[width=4cm]{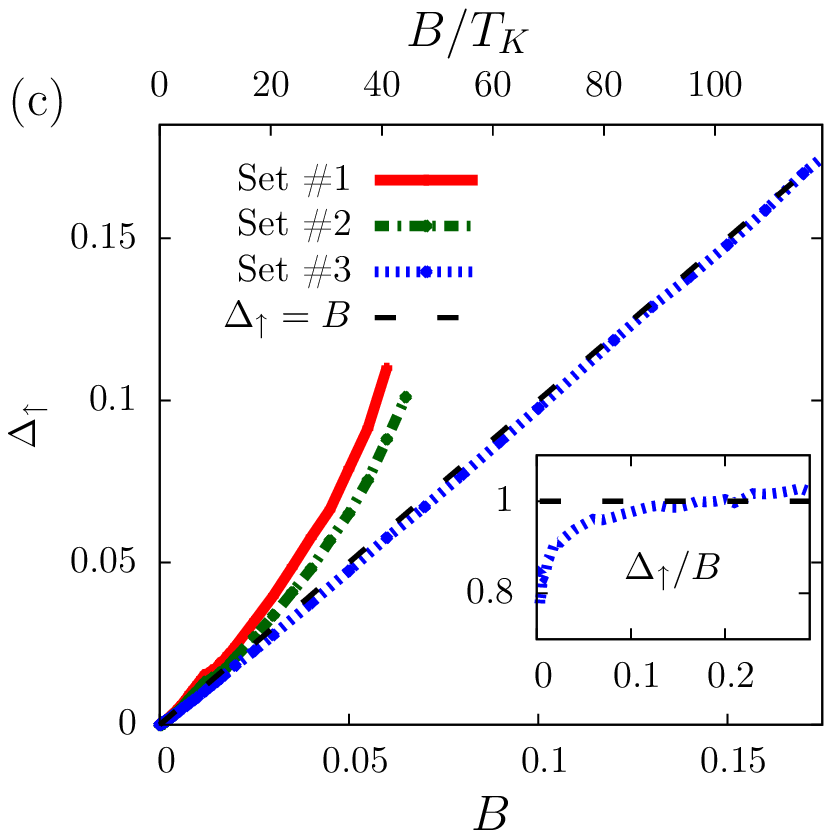}
  \caption{(Color online) (a) Spectral function of the SIAM in a finite magnetic field for three different NRG parameters (see main text).
    The inset displays a close-up of the region around the Fermi energy and the arrows indicate the position of the maximum.  
    (b) Maximum position of the Zeeman-split Kondo resonance in the spin-resolved ($\Delta_\uparrow$) and total spectral function
    ($\delta_\uparrow$) as function of magnetic field. 
    (c) $\Delta_\uparrow$ as function of the magnetic field. The inset shows for set \#3 the position normalized with the magnetic field.
    $\Delta_\uparrow/B=1$ corresponds to the Zeeman splitting.  
    The upper scale on panels (b) and (c) measures the field in units of the Kondo temperature $T_K$. 
}
  \label{fig1}
\end{figure}

We can indeed reproduce the published data\cite{zhangKondoMag10} for 
the SIAM at $T=0$ with $U=-2\epsilon=1$ and $\Gamma=0.16$, which are depicted as set \#1 in Fig.~\ref{fig1}.
For this set the discretization parameter is $\Lambda=2.5$, only a small number of states  $N_s=150$ are retained in
each NRG iteration, a large broadening of $\alpha=0.8$ is used, $N_z=20$ different discretizations are averaged 
($z$-averaging), and  an additional shift\cite{weichselbaumNRG07} $\gamma=\alpha/4$ is employed in the broadening procedure. 
In the following, however, we demonstrate the strong sensitivity of peak maxima on the NRG parameters.
While set \#2 differs from set \#1 only by setting $\gamma=0$,
for set \#3, a discretization parameter $\Lambda=2$ is used, $N_s=1800$ states are kept,
$\alpha=0.075$, and  $N_z=12$ different conduction band 
discretizations are averaged.
All data are calculated within the complete Fock-space algorithm\cite{petersNewNRG06}
also employed by Zhang \etal\cite{zhangKondoMag10}
which coincides with  full density-matrix approach\cite{weichselbaumNRG07}  at $T=0$.

The spin-resolved spectral function $\rho_\uparrow(\omega)$
in a finite magnetic field $B=0.003273\approx 2.2 T_K$ is shown in  Fig.~\ref{fig1}(a) 
for these three different parameter sets. While the results share the same 
qualitative features, they still differ considerably.
The differences do not only affect high energy features like the 
Hubbard satellites, but also the low energy Kondo resonance (see inset). 
The width of the latter is much narrower for set \#3 and, most important
for the present discussion, the positions of the maximum differ (indicated by the arrows 
in the inset).
In panel (b) the positions of the Kondo maxima in total  and spin-resolved 
spectral functions are  compared. For set \#3, $\delta_\uparrow$ and $\Delta_\uparrow$
very quickly approach each other
and are indistinguishable for $B\gtrsim0.004\approx 3T_K$ (see upper scale in the plots). 
In contrast, the curves 
for set \#1  do not approach each other as  observed in  
Ref.~\onlinecite{zhangKondoMag10}.
Panel (c) displays the maximum of the Kondo resonance  
$\Delta_\uparrow$ in $\rho_\uparrow(\omega)$ as function of
the applied magnetic field $B$. The curve of set \#3 increases
almost linearly for all fields. 
It is slightly below the Zeeman splitting and exceeds it 
only for very large fields $B\gtrsim 0.2=1.25\Gamma\approx 150 T_K$ (see inset).
In contrast, the curves for sets \#1 and \#2 increase much faster
and exceed the Zeeman-splitting already for fields $B\gtrsim 0.02\approx 15 T_K$.

The discretization of the conduction band\cite{bullaNRGReview08} within NRG represents a severe approximation 
and recovering the correct continuum limit (e.g.\ $\Lambda\to 1^+$ and $N_s\to\infty$)
is conceptually  far from trivial.
The broadening procedures adopted in the standard
approaches\cite{bullaNRGReview08,freynNRGBroad09} 
offer no concise method,  how to choose specific values for
the parameters involved.
But as these parameters are artificial and remain  $\Lambda$ dependent, the physical phenomenon of interest  
 \textit{must not}  depend on the actual value of these 
parameters.
In Fig.~\ref{fig1} we demonstrate that the results of Zhang \etal\cite{zhangKondoMag10}
are non-universal since they are strongly dependent on the broadening parameters.
Even changing only the artificial shift from $\gamma=\alpha/4$  (as in set \#1) 
to $\gamma=0$ (as in set \#2) considerably changes the peak positions. In contrast,
changing the same parameter for set \#3 (within reasonable bounds, e.g.\ 
$0\leq \gamma <\alpha$) does \textit{not} change
the spectral functions. This also holds for the other parameters, where the artificial 
broadening $ \alpha$ is most critical. Increasing $\alpha$ reduces the critical field,
where the $\Delta_\uparrow$ exceeds the Zeeman splitting.  
Generally, the accurate extraction of Zeeman-split  peak positions from NRG spectral functions 
is a very delicate task as these always occur at finite energies, 
where the discretization errors are in principle not negligible.    


We therefore argue that 
the two results of Zhang \etal\cite{zhangKondoMag10}
stated above are inconclusive and have been obtained 
by a particular choice of NRG parameters. Since there does 
not exist a ``correct'' choice for these parameters,
physical conclusions must only be drawn from NRG spectra 
which are robust to changes of NRG parameters or error
bars should be given. 
Neither requirement appears to be fulfilled by 
Ref.~\cite{zhangKondoMag10} as demonstrated above.

However, we  reproduce the crossover in the Zeeman splitting
from Kondo-like behavior, $\Delta_\uparrow/B <1$, 
to a non-universal splitting with $\Delta_\uparrow/B >1$ [see inset in Fig~\ref{fig1}(c)]. 
But this crossover occurs at  fields  of the order of the charge scale,
$B\sim \Gamma$, that is for fields two orders of magnitude larger than reported
in Ref.~\cite{zhangKondoMag10}.

\section{Acknowledgments}

We thank Hui Zhang for helpful discussions and for sharing the details of his calculations with us.
We acknowledge financial support from the Deutsche Forschungsgemeinschaft under AN  275/6-2.

\bibliography{hui_bibliography}

\end{document}